\documentclass[12pt]{article}

\normalbaselines \topmargin -0.5 truein  \oddsidemargin 0.30
truein \evensidemargin 0.30 truein 
scaled \magstep2  

\usepackage{graphicx}

\newcommand{\nablab}{{\mathop {\rule{0pt}{0pt}{\nabla}}\limits^{\bot}}\rule{0pt}{0pt}}

\begin{document}


\begin{center}
{\bf GRADIENT MODELS OF THE AXION-PHOTON COUPLING}
\end{center}

\vskip 0.8cm


\vskip 0.3cm \centerline{\bf Alexander B. Balakin\footnote{e-mail: Alexander.Balakin@ksu.ru},
Vladimir V. Bochkarev\footnote{e-mail: Vladimir.Bochkarev@ksu.ru}}
\vskip 0.1cm
\centerline{\bf and Nadezhda O. Tarasova\footnote{e-mail: NadezhdaTarasova@yandex.ru}}

\vskip 0.3cm
\centerline{\it Kazan Federal University, Institute of Physics,}

\centerline{\it Kremlevskaya street 18,
420008, Kazan,  Russia}

\vskip 3cm


\noindent
{\bf Abstract}

\noindent
We establish an extended version of the Einstein - Maxwell - axion model by introducing into the Lagrangian cross-terms, which contain
the gradient four-vector of the pseudoscalar (axion) field in convolution with the Maxwell tensor. The gradient model of the axion-photon coupling is applied to cosmology:
we analyze the Bianchi-I type Universe with an initial magnetic field, electric field induced by the axion-photon interaction, cosmological constant and dark matter, which is described in terms of the pseudoscalar (axion) field. Analytical, qualitative and numerical results are presented in detail for two distinguished epochs: first, for the early Universe with magnetic field domination; second, for the stage of late-time accelerated expansion.

\newpage

\section{Introduction}
\label{intro}

Basic elements of the modern theory of the axion - photon (pseudoscalar-photon) coupling have been formulated thirty four years ago, in 1977-1978.
Peccei and Quinn \cite{Peccei0} in the context of the strong CP- problem have opened the discussion about new pseudoscalar particles, which could appear as a result of
a spontaneous phase transition (PQ- symmetry breakdown in modern terminology). Due to the works of Weinberg \cite{Weinberg0} and Wilczek \cite{Wilczek0} now we associate this problem
with the theory of pseudo-Goldstone bosons and with the so-called WIMPs (Weakly Interacting Massive Particles). For particles of this type the term axions was introduced by Wilczek;
one can find the history of axions, e.g., in \cite{Turner}-\cite{Battesti}. At the same time, in 1977, studying the Equivalence Principle, Ni \cite{Ni0I}  suggested to describe
the pseudoscalar - photon interactions on the electrodynamic language: the cross-invariant $\phi F^{*}_{mn}F^{mn}$ was introduced into the Lagrangian, where $\phi$, the dimensionless pseudoscalar field, appears in the product with the convolution of Maxwell tensor $F^{mn}$ and its dual $F^{*}_{mn}$. Later, on the basis of these ideas, new theoretical models were established and developed, namely, axion electrodynamics (see, e.g., \cite{Wilczek} -\cite{Ni11} ) and its covariant generalizations: the Einstein - Maxwell - (dilaton) - axion models (see, e.g., \cite{Claus1} -\cite{Matos} for review and references). Nowadays, the axion electrodynamics provides a theoretical base for the experimental search for the axion-photon coupling (see, e.g., \cite{Zavattini1} - \cite{Ni25}).
The most intriguing application of the axion theory is connected with the concept of dark matter, whose contribution into the
Universe energy balance is estimated to be about 23 \% (see, e.g., \cite{Raffelt,Battesti,Shellard,Duffy,Sikivie,Visinelli}. If the hypothesis, that the dark matter is composed of axions, will be proved, and about 23\% of the Universe energy will be prescribed to the contribution from the so-called background (relic) axions, we could conclude that the axion-photon coupling is the important element in the hierarchy of cosmic interactions.

 By analogy with classical Maxwell-Faraday electrodynamics, the axion electrodynamics admits generalizations of different types. There are various reasons for the extension of this theory; for instance, the nonminimal extension of the Einstein - Maxwell - axion theory (see, \cite{BaWTNi,BaWTNi2}) was motivated, in particular, by the problems of correct estimation of the polarization rotation effect for the electromagnetic waves running through the regions with nonstationary and/or nonuniform gravitational field. Other arguments for the extension of the axion electrodynamics come from the analogy with  electromagnetic theory of moving anisotropic inhomogeneous (polarizable, magnetizable, rheological) media \cite{EM,LLP,HehlObukhov}: in these models the tensor of linear response depends on the macroscopic velocity of the medium as a whole, and its derivatives (see, e.g., \cite{Ba07,Ba071}), thus providing the axion system to possess the birefringence, dynamo-optical, etc. effects.

The gradient extension of the Einstein - Maxwell - axion model presented in this work has the following argumentation. First of all,
when the gradient four-vector of the pseudoscalar field, $\nabla_i \phi$, is nonvanishing, the axion - photon system can be considered as an anisotropic one - axis quasimedium, with a director $\Re_i$, normalized four-vector, which is proportional to $\nabla_i \phi$. Anisotropy of continua is expected to be inherited by permittivity tensors \cite{LLP}, thus providing the birefringence effect. However,  the standard axion electrodynamics with the cross-invariant $\phi F^{*}_{mn}F^{mn}$ in the Lagrangian does not take into account the anisotropy of the constitutive equations, and  this weak point of theory can be amended by the gradient type extension.
The second motive to insert the gradient four-vector $\nabla_i \phi$ into the Lagrangian comes from the analogy with the so-called derivative coupling in the nonminimal theory of the scalar fields \cite{derc1,derc2,derc3} and of the Higgs multiplets \cite{BZD1,BZD2,BZD3}. Let us remind that in \cite{derc1,derc2,derc3} in addition to the standard invariant $\xi R \Phi^2$ with Ricci scalar $R$ and scalar field $\Phi$, the new term $\alpha R^{ik} \nabla_i \Phi \nabla_k \Phi$, with the Ricci tensor $R^{ik}$, was inserted into the Lagrangian. In our model in addition to the term  $\phi F^{*}_{mn}F^{mn}$ we introduce admissible cross-invariants, which contain the convolution $F^{*}_{ik} \nabla^k \phi$. The third argument to extend the theory of axion-photon coupling is connected with the attempts to detect experimentally the birefringence in the vacuum magnetic field (see, e.g., \cite{Zavattini1} - \cite{Ni25}): the standard vacuum axion electrodynamics does not admit birefringence, while its gradient-type modifications permit it.

The paper is organized as follows. In Sec.~\ref{sec2} based on the Lagrange formalism we obtain self-consistent set of master equations for the electromagnetic (Sec.~\ref{subsec22}), pseudoscalar (Sec.~\ref{subsec23}), gravitational fields (Sec.~\ref{subsec24}), and discuss general properties of the model. In Sec.~\ref{subsec25} we discuss the problem of bilingual description of the axion-photon interactions, and constraint the choice of the phenomenological parameters of the gradient model using two exact solutions for the induced electric field.   In Sec.~\ref{sec3} we consider obtained master equations by the example of anisotropic cosmological model with initial magnetic field, axionically induced electric field, dark energy in $\Lambda$-term representation and dark matter described in terms of pseudoscalar (axion) field. The reduced system of master equations (Sec.~\ref{subsec31}) is presented in terms of associated dynamic system (Sec.~\ref{subsec33}). The epochs of dark matter domination (Sec.~\ref{subsec34}) and of magnetic field domination (Sec.~\ref{subsec35}) are distinguished and studied analytically and qualitatively; the model as a whole is analyzed numerically.

\section{Two-parameter gradient-type model \\ of the axion-photon coupling}
\label{sec2}

\subsection{General formalism}
\label{subsec21}

The gradient-type extension of the axionic Einstein-Maxwell model is based on the Lagrangian formalism, the action functional under consideration is of the form
$$
S {=} \int d^4 x \sqrt{{-}g} \left\{ \frac {R{+}2\Lambda}{\kappa}{+}\frac{1}{2}F_{mn}F^{mn} {+}\frac{1}{2}\phi F^{*}_{mn} F^{mn}
{+} \Psi_{0}^2\left[ {-}\nabla_m\phi \nabla^m\phi {+}V(\phi^2) \right] {+}
\right.
$$
\begin{equation}
\left.
{+}\frac{1}{2}\lambda_1 F_{mn}F^{mn} \nabla^p \phi \nabla_p \phi {+} \frac{1}{2} \lambda_2 F_{mp} F^{mq} \nabla^p \phi \nabla_q \phi \right\} \,.
\label{action}
\end{equation}
Here $g$ is the determinant of the metric tensor $g_{ik}$,
$\nabla_{m}$ is a covariant derivative, $R$ is the Ricci scalar, $\kappa{=}\frac{8\pi G}{c^4}$ is the Einstein coupling constant, $\Lambda$ is the cosmological constant.
The Maxwell tensor $F_{mn}$ is given by
\begin{equation}
F_{mn} \equiv \nabla_m A_{n} {-} \nabla_n A_{m} \equiv 2\nabla_{[m} A_{n]}\,,
\label{maxtensor}
\end{equation}
where $A_m$ is an electromagnetic potential four-vector; $F^{*mn}
{=} \frac{1}{2} \epsilon^{mnpq}F_{pq}$ is the tensor dual to
$F_{pq}$; $\epsilon^{mnpq} = \frac{1}{\sqrt{{-}g}} E^{mnpq}$ is
the Levi-Civita tensor, $E^{mnpq}$ is the absolutely antisymmetric
Levi-Civita symbol with $E^{0123}{=}1$. The dual Maxwell tensor
satisfies the condition
\begin{equation}
\nabla_{k} F^{*ik} =0 \,. \label{Emaxstar}
\end{equation}
The first term $\frac{R}{\kappa}$ corresponds to the standard Hilbert - Einstein Lagrangian;
the second term is the standard Lagrangian for an electromagnetic
field; the third term is the pseudoscalar - photon interaction
Lagrangian; the fourth and fifth terms
constitute the pseudoscalar Lagrangian.
The symbol $\phi$ stands for a pseudoscalar field, this quantity
being dimensionless. The axion field itself, $\Phi$, is considered to
be proportional to this quantity $\Phi = \Psi_0 \phi$ with a
constant $\Psi_0$. The function $V(\phi^2)$ describes the potential of the pseudoscalar field.
Two last terms in the Lagrangian introduce new terms containing the gradients of the pseudoscalar field in the product and/or in convolution with the Maxwell tensor;
the parameters $\lambda_1$ and $\lambda_2$ are phenomenological coupling constants.

\subsection{Axion electrodynamics}
\label{subsec22}

Variation of the action functional (\ref{action}) with respect
to the four-vector potential $A_i$ gives the equations of axion electrodynamics
\begin{equation}
\nabla_k H^{ik}=0 \,,
\label{eld1}
\end{equation}
where the excitation tensor $H^{ik}$ is given by the term
\begin{equation}
H^{ik}{=} F^{ik} {+} \phi F^{*ik} {+} \lambda_1  F^{ik} \nabla_q \phi \nabla^q \phi
{+} \lambda_2 \nabla^{[k} \phi F^{i]q} \nabla_q \phi .
\label{eld2}
\end{equation}
Using the linear constitutive equations
\begin{equation}
H^{ik} = C^{ikmn} F_{mn} \,,
\label{eld3}
\end{equation}
we readily obtain that the linear response tensor $C^{ikmn}$ takes now the form
$$
C^{ikmn} = \frac{1}{2} \left( g^{im} g^{kn} {-} g^{in} g^{km} \right) \left(1 {+}
\lambda_1 \nabla_p \phi \nabla^p \phi \right) {+}
$$
\begin{equation}
{+}\frac{1}{2} \phi \epsilon^{ikmn} {+}
\frac{\lambda_2}{2} \left( g^{i[m} \nabla^{n]} \phi \nabla^k \phi
{+} g^{k[n} \nabla^{m]} \phi \nabla^i \phi \right) \,.
\label{eld4}
\end{equation}
It is well known that using the medium velocity four-vector $U^i$, normalized by
$U^iU_i=1$, one can decompose the linear response tensor $C^{ikmn}$ uniquely as
$$
C^{ikmn} = \varepsilon^{i[m} U^{n]} U^k + \varepsilon^{k[n} U^{m]} U^i +
$$
\begin{equation}
-\frac12 \eta^{ikl}(\mu^{-1})_{ls}  \eta^{mns} + \eta^{ikl} U^{[m}\nu_{l}^{\
n]} + \eta^{lmn} U^{[i} \nu_{l}^{\ k]} \,. \label{eld5}
\end{equation}
Here $\varepsilon^{im}$ is the dielectric permittivity tensor,
$(\mu^{-1})_{pq}$ is the magnetic impermeability tensor, and
$\nu_{p \ \cdot}^{\ m}$ is the tensor of magneto-electric
coefficients. These quantities are defined as follows
\begin{eqnarray}
\varepsilon^{im} &=& 2 C^{ikmn} U_k U_n\, \nonumber\\
(\mu^{-1})_{pq}  &=& - \frac{1}{2} \eta_{pik} C^{ikmn}
\eta_{mnq}\,,
\nonumber\\
\nu_{p}^{\ m} &=& \eta_{pik} C^{ikmn} U_n =U_k C^{mkln} \eta_{lnp}\,. \label{eld6}
\end{eqnarray}
The tensors $\eta_{mnl}$ and $\eta^{ikl}$ are the anti-symmetric
ones orthogonal to $U^i$ and defined as
\begin{equation}
\eta_{mnl} \equiv \epsilon_{mnls} U^s \,,
\quad
\eta^{ikl} \equiv \epsilon^{ikls} U_s \,.
\label{eld7}
\end{equation}
They are connected by the useful identity
\begin{equation}
- \eta^{ikp} \eta_{mnp} = \delta^{ikl}_{mns} U_l U^s = \Delta^i_m
\Delta^k_n - \Delta^i_n \Delta^k_m \,, \label{eld8}
\end{equation}
where the symmetric projection tensor $\Delta^{ik}$ is
\begin{equation}
\Delta^{ik} = g^{ik} - U^i U^k \,.
\label{eld9}
\end{equation}
The tensors $\varepsilon_{ik}$ and $(\mu^{-1})_{ik}$ are
symmetric, but $\nu_{lk}$ is in general non-symmetric. Clearly, these three
tensors are orthogonal to $U^i$.
Using the expression (\ref{eld4}) one can calculate tensors
$\varepsilon^{im}$, $(\mu^{-1})_{im}$ and $\nu^{pm}$ explicitly, yielding
\begin{equation}
\varepsilon^{im} = \Delta^{im}\left[1{+} \lambda_1 \nabla_q \phi \nabla^q \phi \right] +
\frac{1}{2}\lambda_2 \left[ \Delta^{im} \left( \cal D \phi \right)^2 {+} \nablab^i \phi \nablab^m \phi \right]\,,
\label{eld11}
\end{equation}
\begin{equation}
\left(\mu^{-1}\right)_{im} = \Delta_{im} \left[ 1+ \lambda_1 \nabla_q \phi \nabla^q \phi \right]+
\frac{1}{2}\lambda_2 \left[\Delta_{im}\nablab_q \phi \nablab^q \phi - \nablab_i \phi \nablab_m \phi \right] \,,
\label{eld12}
\end {equation}
\begin{equation}
\nu^{pm} = - \phi \Delta^{pm} + \frac{1}{2} \lambda_2 {\cal D}\phi \ \eta^{pmk}
\nablab_k \phi \,.
\label{eld13}
\end{equation}
The symmetric tensors $\varepsilon^{im}$ and $(\mu^{-1})_{im}$
contain both new coupling parameters $\lambda_1$ and $\lambda_2$, while the
cross-tensor $\nu_{lk}$, which describes optical activity effects (see, e.g., \cite{nu1}),
contains the second coupling parameter only. Here we used the standard definitions
\begin{equation}
{\cal D} \equiv U^m \nabla_m , \quad \nablab_k \equiv \Delta^{m}_{k} \nabla_m \,.
\label{eld14}
\end{equation}
for the convective derivative and pure spatial gradient, respectively.

One can mention two interesting details. First, the tensors
$\varepsilon^{im}$ and  $(\mu^{-1})_{im}$ become spatially anisotropic, when the pseudoscalar field is inhomogeneous and $\lambda_2 \neq 0$;
this means that the axion-photon interaction of the gradient type can produce the phenomenon of birefringence in the course of electromagnetic wave propagation.
Second, the cross-tensor $\nu^{pm}$ contains both symmetric  and skew-symmetric terms (see the terms
with $\Delta^{pm}$ and $\eta^{pmk}$, respectively). It is well-known that the non-vanishing cross-tensor
$\nu^{ik}$ indicates that the axionic vacuum as a (quasi) medium is optically active, and the
rotation of the Faraday type takes place in the course of
electromagnetic wave propagation. Thus, one can see directly from
(\ref{eld13}), that the interaction of a new gradient type between
electromagnetic and axion fields produces optical activity of new type, if the pseudoscalar field is non-stationary, i.e., ${\cal D} \phi \neq 0$, and inhomogeneous, i.e.,
$\nablab_k \phi \neq 0$, simultaneously.

\subsection{Pseudoscalar field evolution}
\label{subsec23}

Variation of the action functional (\ref{action}) with respect to the pseudoscalar field $\phi$ gives the equation
\begin{equation}
\nabla_q \left[ \left(g^{pq}  {-} \Theta^{pq}\right) \nabla_p \phi \right] {+} \phi V' \left( \phi^2 \right)=
{-} \frac{1}{4 \Psi_{0}^2 } F^{*}_{mn} F^{mn} \,,
\label{ps1}
\end{equation}
where the tensor $\Theta^{pq}$ is defined as
\begin{equation}
\Theta^{pq} = \frac{1}{2\Psi_{0}^2 } \left[\lambda_1 g^{pq} F_{mn} F^{mn} {+} \lambda_2 F^{mp} F_{m}^{\ q} \right]\,,
\label{ps2}
\end{equation}
and the prime denotes the derivative with respect to the argument $\phi^2$. The pseudoscalar $F^{*}_{mn} F^{mn}$ plays the role of the local source of the axion field,
as in the standard theory of the axion-photon interaction. The novelty of the model is connected with the tensor
\begin{equation}
\tilde{g}^{pq} = \left(g^{pq}  - \Theta^{pq}\right) \,,
\label{ps3}
\end{equation}
which in fact plays the role of an effective metric for the pseudoscalar waves: these waves propagate in axionic (quasi) vacuum similarly to the waves in the medium, so that the
velocity of pseudoscalar waves depends on the structure of the electromagnetic field (see, e.g., \cite{BZD1,BZD2,BZD3}, in which color-acoustic analogs of these waves have been studied).

\subsection{Gravity field equations}
\label{subsec24}

Modified Einstein's equations obtained by the variation of the action functional (\ref{action}) with respect to the metric $g^{pq}$ can be presented in the form
\begin{equation}
R_{pq} {-} \frac{1}{2} R g_{pq}  = \Lambda g_{pq} + \kappa \left[T_{pq}^{(EM)} + T_{pq}^{(A)} + \lambda_1 T_{pq}^{(1)} + \lambda_2 T_{pq}^{(2)} \right] \,.
\label{gr1}
\end{equation}
The stress-energy tensor of the electromagnetic field
\begin{equation}
T_{pq}^{(EM)} = \frac{1}{4} g_{pq} F_{mn} F^{mn} - F_{pm} F_{q}^{\ m}
\label{gr2}
\end{equation}
and the stress-energy tensor of the pure axionic field
\begin{equation}
T_{pq}^{(A)} {=}\Psi_{0}^2 \left\{ \nabla_p \phi \nabla_q \phi {-}  \frac{1}{2} g_{pq} \left[ \nabla_m \phi \nabla^m \phi
{-} V\left( \phi^2 \right) \right] \right\}
\label{gr3}
\end{equation}
are presented by the well-known terms. Two terms
\begin{equation}
T_{pq}^{(1)} = - \frac{1}{2} F_{mn}F^{mn} \nabla_p \phi \nabla_q \phi + T_{pq}^{(EM)} \nabla_n \phi \nabla^n
\phi \,,
\label{gr4}
\end {equation}
and
\begin{equation}
T_{pq}^{(2)} = \frac{1}{4} g_{pq} F_{m}^{\ n} F^{ml} \nabla_n \phi \nabla_l \phi {-} \frac{1}{2}\nabla^l \phi \left[F_{ql} F_{p}^{\ m}
  \nabla_m \phi  {+} F_{ml} F^{m}_{\ q}  \nabla_p \phi {+} F_{ml} F^{m}_{\ p}  \nabla_q \phi \right] \,,
\label{gr5}
\end{equation}
describe new source-terms in the right-hand side of the gravity field equations.

Thus, the complete self-consistent system of master equations describing the extended Einstein - Maxwell - axion model is derived: it contains the electrodynamic equations (see (\ref{eld1}),
(\ref{eld2})), (\ref{Emaxstar}), the equation for the pseudoscalar (axion) field (see (\ref{ps1}), (\ref{ps2}))  and the gravity field equations (\ref{gr1})-(\ref{gr5}).

\subsection{Dark matter and alternative description \\ of the axionic source-term}
\label{subsec25}

Axions are considered to be candidates for the cosmic substrate indicated as dark matter (see, e.g., \cite{Battesti,Shellard,Duffy,Sikivie}). There are two different approaches to the description of axions:
the first one is based on the pseudoscalar field theory, the second approach uses the terminology of the theory of continuous media. In the first approach the stress-energy tensor of the axions is quadratic in the gradient four-vector $\nabla_k \phi$ and is the direct result of the variation procedure. In the second case we deal with some phenomenological representation, and the best known construction is the stress-energy tensor of an imperfect fluid with the energy density $W$, heat flux four-vector $q^i$ and the (anisotropic) pressure tensor ${\cal P}_{ik}$. Let us assume that these quantities coincide, i.e.,
$$
\Psi^2_0 \left\{\nabla_i \phi \nabla_k \phi {-}
\frac{1}{2} g_{ik} \left[ \nabla^m \phi \nabla_m \phi {-}
V(\phi^2) \right] \right\} =
$$
\begin{equation}
\hspace{10mm} = W U_i U_k + q_i U_k + q_k U_i + {\cal P}_{ik} = T_{ik}^{({\rm A})}\,.
\label{cl12}
\end{equation}
Clearly, the compatibility of such equality is connected with some physical and mathematical constraints. Let us start with the discussion of the heat flux four-vector. Usually, the axionic systems,
considered as a fluid, are characterized by vanishing heat flux four-vector
\begin{equation}
q^i \equiv  U^p T_{pq}^{({\rm A})} \Delta^{iq} = 0\,.
\label{q1}
\end{equation}
Mathematically, this can be obtained when the macroscopic velocity four-vector $U^i$ is the eigen-vector of the stress-energy tensor (the Landau-Lifshitz definition). Direct calculation of $q^i$ using the left-hand side of the equality (\ref{cl12}) yields $q^i{=} \Psi^2_0 {\cal D}\phi \nablab^i \phi{=}0$. When the pseudoscalar field is homogeneous, i.e., $\nablab^i \phi{=}0$, or when the pseudoscalar field is stationary, i.e., ${\cal D}\phi {=}0$, one obtains readily that $q^i{=}0$. To provide the equality $q^i{=}0$ in general case (inhomogeneous and non-stationary simultaneously) one can assume that the macroscopic velocity four-vector is chosen to be orthogonal to the gradient four - vector of the pseudoscalar field, i.e.,  ${\cal D}\phi {=} U^i \nabla_i \phi{=}0$. This equation can be generally resolved, when $\nabla_k \phi$ is a spacelike four-vector.
The energy-density scalar $W$ can be written as follows
\begin{equation}
W \equiv U^p T_{pq}^{({\rm A})}U^q = \frac{1}{2} \Psi^2_0 \left[({\cal D}\phi)^2 + V - \nablab^i \phi \nablab_i \phi \right]\,.
\label{cl13}
\end{equation}
 This quantity is always positively defined. The structure of the pressure tensor also attracts special attention. In many works the dark matter is considered to have the isotropic (Pascal) pressure $P$, moreover, in the cold dark matter model this quantity is vanishing $P \to 0$. The effective pressure tensor presented in terms of derivatives of the pseudoscalar field has in general case both isotropic and anisotropic parts
$$
 {\cal P}_{ik} \equiv \Delta^p_i T_{pq}^{({\rm A})}\Delta^q_k =
 $$
 \begin{equation}
 = \frac{1}{2} \Psi^2_0 \Delta_{ik} \left[V{-}({\cal D}\phi)^2 {-} \nablab^m \phi \nablab_m \phi \right] + \Psi^2_0 \nablab_i \phi \nablab_k \phi \,.
\label{cl14}
\end{equation}
Clearly, when the pseudoscalar field is homogeneous, the effective pressure tensor is spatially isotropic. Thus, keeping in mind the bilingual description of dark matter axions in terms of perfect fluid and in terms of pseudoscalar field, we have to provide two constraints on the pseudoscalar field model: first, the heat flux to be vanishing, second, the pressure tensor to be isotropic. Below we remind two explicit examples as illustrations that such models exist.

\subsubsection{First explicit example: spatially homogenous cosmological models}
\label{subsubsec251}

Anisotropic spatially homogeneous cosmological model of the Bianchi-I type is known to be characterized by the metric
\begin{equation}
ds^2 = dt^2 {-} a^2(t) \ (dx^1)^2 {-} b^2(t) \ (dx^2)^2 {-} c^2(t) \
(dx^3)^2 \,. \label{metric}
\end{equation}
In the framework of spatially homogeneous Bianchi-I cosmological models, as well as for the isotropic FLRW models obtained by $a{=}b{=}c$, one assumes that all functions of state depend on time only, so that $\nablab_i \phi {=}0$. The macroscopic velocity four-vector is of the form $U^i{=}\delta^i_0$, and thus ${\cal D}\phi = \dot{\phi}$, where $\dot{\phi}$ is the time derivative of the pseudoscalar (axion) field. In these models the heat flux four-vector vanishes. We use here the well-known procedure (see, e.g., \cite{NO} for details): the energy density of the dark matter can be expressed as follows
\begin{equation}
\frac{1}{2}\Psi^2_0 \left[\dot{\phi}^2 + V(\phi^2) \right] = W(t) \,.
\label{cl15}
\end{equation}
Clearly, the pressure tensor (\ref{cl14}) is now isotropic
\begin{equation}
{\cal P}_{ik} = - \Delta_{ik} P \,,
\label{cl16}
\end{equation}
where $P$ is the Pascal (isotropic) pressure of the dark matter
\begin{equation}
\frac{1}{2}\Psi^2_0 \left[\dot{\phi}^2 - V(\phi^2)\right] = P(t) \,.
\label{cl17}
\end{equation}
These two relations yield the time derivative
\begin{equation}
\frac{d \phi}{dt} = \pm \frac{1}{\Psi_0} \sqrt{W(t)+P(t)} \,,
\label{cl18}
\end{equation}
and the pseudoscalar field itself
\begin{equation}
\phi(t) - \phi(t_0)= \pm \frac{1}{\Psi_0} \int^{t}_{t_0} dt' \sqrt{W(t')+P(t')}  \,,
\label{cl19}
\end{equation}
in terms of dark matter energy density and pressure. The pseudoscalar field potential $V(\phi^2)$ can be reconstructed using the relation
\begin{equation}
V(\phi^2) = \frac{1}{\Psi^2_0} \left[W(t) - P(t) \right]
\label{cl20}
\end{equation}
and the function $\phi(t)$ obtained in (\ref{cl19}).

Let us mention that in present epoch of the Universe accelerated expansion the ground state of the dark matter is considered to be non-relativistic, i.e., $P<<W$.
More precisely, the CMD model (Cold Dark Matter) assumes that $P{=}0$ and one deals with a dust. In this case using (\ref{cl17}) one can conclude that
$\frac{1}{2}\dot{\phi}^2{=}\frac{1}{2}V(\phi^2)$ in line with the well-known classical virial theorem. Thus, in order to obtain the CDM model one should put $P{=}0$
into (\ref{cl18})-(\ref{cl20}), and replace the total energy density $W(t)$ by the rest energy density $\rho c^2$, where $\rho(t)$ is the mass density of the dark matter dust.

The dielectric permittivity tensor can be now reduced to the scalar function $\varepsilon(t)$
\begin{equation}
\varepsilon^{im} {=} \Delta^{im} \varepsilon(t) \,, \quad \varepsilon(t)= 1 + \left(\lambda_1+ \frac{1}{2}\lambda_2 \right) \dot{\phi}^2 \,.
\label{ueld11}
\end{equation}
Similarly, the magnetic impermeability tensor has the form
\begin{equation}
\left(\mu^{-1}\right)_{im} = \Delta_{im} \frac{1}{\mu(t)} \,, \quad
\frac{1}{\mu(t)} = 1{+} \lambda_1 \dot{\phi}^2 \,.
\label{ueld12}
\end {equation}
The cross-tensor is now pure symmetric
\begin{equation}
\nu^{pm} = - \phi \Delta^{pm} \,.
\label{ueld13}
\end{equation}
The tensors $\varepsilon^{im}$ and $\left(\mu^{{-}1}\right)_{im}$ are spatially isotropic, thus, the birefringence effect is absent, if we restrict ourselves by spatially homogeneous cosmological models.
The refraction index $n(t)$ can be introduced by the relation
\begin{equation}
n^2(t) = \varepsilon(t) \mu(t) = \frac{1 + \left(\lambda_1+ \frac{1}{2}\lambda_2 \right) \dot{\phi}^2}{1{+} \lambda_1 \dot{\phi}^2} \,.
\label{ueld139}
\end{equation}
When $\lambda_1{=}\lambda_2{=}0$, one obtains that $n^2{=}1$, the phase velocity of electromagnetic waves $v_{{\rm ph}}{=}\frac{c}{n(t)}$ coincides with speed of light in vacuum. In general case the phase velocity depends on time through the function $\dot{\phi}^2$. The phase velocity can take infinite value, when the denominator of (\ref{ueld139}) vanishes; it is equal to zero, when the numerator of (\ref{ueld139}) vanishes. The function $n^2(t)$ can be (in principle) negative, and in this (unlighted) epoch the electromagnetic waves do not propagate in the Universe, do not scan its internal structure and can not bring information to observers.

The Bianchi-I model with pure magnetic field is well-studied (see, e.g., \cite{ExactSolutions,dina} for review). The novelty of our model is that due to the axion-photon coupling the magnetic field $B^{3}(t) {=} {-} \frac{1}{abc}F_{12}\neq 0$ produces the electric field, so we consider the electromagnetic configuration with supplementary component $E^3{=} F^{30}(t) \neq 0$. When all the quantities depend on time only, the equations (\ref{Emaxstar}) are satisfied, if  $F_{12}{=}const$. The equations (\ref{eld1}) with the excitation tensor (\ref{eld2}) give the following exact solution for the electric field
\begin{equation}
F^{30}(t) {=} \frac{F_{12} \phi(t) + const}{a(t)b(t)c(t)\left[1 {+} \left(\lambda_1 {+} \frac{1}{2} \lambda_2\right) \dot{\phi}^2(t) \right]}
\,.
\label{cosm301}
\end{equation}
Clearly, this electric field can reach infinite value, if the parameter $\lambda_1 {+} \frac{1}{2} \lambda_2$ is negative; this means that at some critical moment the invariants of the electromagnetic field can be singular. Thus, excluding the singularity of this type we require that the coupling constants satisfy the inequality  $\lambda_1 {+} \frac{1}{2} \lambda_2 \geq 0$.

\subsubsection{Second explicit example: static models with spherical symmetry}
\label{subsubsec252}

Static spherically symmetric configurations can be cha\-racterized by the metric
\begin{equation}
ds^2 = \sigma^2 N dt^2 {-} \frac{1}{N} dr^2 - r^2 \left(d\theta^2 + \sin^2{\theta}d\varphi^2 \right) \,, \label{metric2}
\end{equation}
where the metric functions $\sigma(r)$ and $N(r)$ depend on the radial variable $r$ only. Assuming that $\phi$ also depends on $r$ only, we obtain readily that
$q^i=0$, and
\begin{equation}
W(r)=\frac{1}{2}\Psi^2_0 \left[V(\phi^2)+N \phi^{\prime 2} \right]
\,, \label{sss1}
\end{equation}
\begin{equation}
-P_{\bot} \equiv {\cal P}^{\theta}_{\theta} = {\cal P}^{\varphi}_{\varphi} = W(r) \,, \label{sss22}
\end{equation}
\begin{equation}
-P_{||} \equiv {\cal P}^{r}_{r} = \frac{1}{2}\Psi^2_0 \left[V(\phi^2)- N \phi^{\prime 2} \right] \,. \label{sss2}
\end{equation}
The prime denotes the derivative with respect to $r$. The profile of the pseudoscalar field $\phi(r)$
can be now reconstructed using the relation
\begin{equation}
\phi^{\prime } = \pm \frac{1}{\Psi_0 \sqrt{N}} \sqrt{W(r)+P_{||}(r)} \,. \label{sss3}
\end{equation}
In order to find constraints for the coupling parameters $\lambda_1$ and $\lambda_2$, let us consider the model with magnetic monopole and radial electric field induced by axion-photon
interactions. The Maxwell equations admit now the following solution
\begin{equation}
F_{\theta \varphi} = \mu \sin{\theta} \,, \quad F^{0r}= \frac{\mu \phi + const}{\sigma r^2 \left[1{-} N\phi^{\prime 2}\left(\lambda_1 {+} \frac{1}{2}\lambda_2\right)\right]} \,. \label{sss4}
\end{equation}
In order to guarantee that the solution is nonsingular for arbitrary pseudoscalar field we should assume that $\lambda_1 {+} \frac{1}{2} \lambda_2 \leq 0$. Combining two constraints obtained for cosmological model and for this static model we see that only the requirement $\lambda_1 {+} \frac{1}{2} \lambda_2 {=} 0$ is appropriate. Below we consider the one-parameter models with $\lambda_2{=}{-}2\lambda_1$.

\section{Cosmological application: \\ the toy model of the Bianchi-I type}
\label{sec3}

The gradient-type extension of the Einstein - Maxwell - axion model formulated here attracts attention, since several new tests can be suggested on the base of this model.
In this paper we consider shortly one cosmological application based on the anisotropic Bianchi-I type model. Our choice can be argued as follows. On the one hand, the model of anisotropic
expanding Universe with a strong magnetic field and axionic dark matter is the best theoretical model for the study of the axion-photon coupling and its consequences.
On the other hand, the impressive detailed analysis of the recent observations of the Cosmic Microwave Background (CMB) radiation (see, e.g., \cite{7Y}) gives a unique possibility to fit
free parameters of the model under consideration and to check the main consequences. Our strategy is the following: in this paper we clarify the roles of guiding parameters of the model
and illustrate (analytically, qualitatively and numerically) basic tendencies in the evolution of main state functions. Next paper, entirely devoted to the fitting of the guiding parameters
of the model and based on the comparison with results of \cite{7Y}, is now in preparation.

Let us define concretely main problems, which we intend to consider in this cosmological application of the established model. {\it First} of all, we intend to demonstrate that the axion-photon
coupling in presence of strong cosmological magnetic field can be a source of a substantial growth of the pseudoscalar (axion) field $\phi$; keeping in mind the hypothesis that the background (relic) axions form the dark matter, we hope to explain large contribution from this cosmic substrate into the total energy-density (23\% in our late-time epoch). {\it Second}, since the axions provide spatially isotropic source of the gravitational field, we intend to show that the substantial growth of the number of axions, caused by the axion-photon coupling, could accelerate the isotropization process of the Universe. {\it Third}, we hope to illustrate the idea that the axion-photon coupling can be a reason for multi-stage inflationary-type regimes of evolution of the early Universe.


Anisotropic cosmological Bianchi-I model characterized by the metric
(\ref{metric}) is assumed to be constructed using three constituents: the dark energy (it is presented by the $\Lambda$-term), the dark matter (it is described by the axions in terms of pseudoscalar field) and the electromagnetic field (it has originally the magnetic component, but the electric component appears due to the axion-photon coupling).
The total set of master equations can be reduced now as follows.

\subsection{Reduced master equations}
\label{subsec31}

\subsubsection{Exact solution to the electrodynamic equations}
\label{subsubsec311}

Due to the axion-photon coupling the magnetic field $B^{3}(t) {=} {-} \frac{1}{abc}F_{12}\neq 0$ ($F_{12}{=}const$) produces the electric field $E^3{=} F^{30}(t) \neq 0$. We consider the model with $\lambda_1{+}\frac{1}{2}\lambda_2{=}0$, thus the equations (\ref{eld1}) with the excitation tensor (\ref{eld2}) give the following exact solution for the electric field
\begin{equation}
F^{30}(t) =  \frac{F_{12} \phi(t)}{a(t)b(t)c(t)} \,.
\label{cosm5}
\end{equation}
The corresponding physical components of the magnetic $B(t)$ and electric $E(t)$ fields, given by
\begin{equation}
B(t) \equiv \sqrt{F_{12}F^{12}}= \frac{F_{12}}{a(t)b(t)} \,,
\label{cosm555}
\end{equation}
\begin{equation}
E(t) \equiv \sqrt{-F_{30}F^{30}}= \frac{F_{12} \phi(t)}{a(t)b(t)} = B(t)\phi(t)
\,,
\label{cosm556}
\end{equation}
are proportional to the constant $F_{12}$, which we will consider as a guiding parameter of the model.

\subsubsection{Reduced equation for the pseudoscalar field}
\label{subsubsec312}

The axion field function $\phi(t)$ can be found from the equation
\begin{equation}
\frac{1}{abc}\frac{d}{dt}\left[ abc \dot{\phi} \left(\Psi^2_0 {-}  \frac{\lambda_1 F^2_{12}}{a^2 b^2} \right)\right] {=} \phi
\left(\frac{ F^2_{12}}{a^2 b^2} {-} \Psi^2_0 V^{\prime}\right) \,.
\label{cosm9}
\end{equation}
When $\lambda_1$ is positive, there exists a moment $t_{({\rm crit})}$, for which the coefficient in parentheses in the left-hand side of (\ref{cosm9}) takes zero value, and the equation becomes degenerated.
In order to guarantee the absence of singularity in the equation (\ref{cosm9}), we assume below that $\lambda_1 \leq 0$ and thus $\lambda_2 {=} {-}2\lambda_1 \geq 0$.

\subsubsection{Reduced Einstein equations}
\label{subsubsec313}

Evolution of the gravitational field is described now by the following system of nonlinear equations:
\begin{equation}
\frac{\dot{a}}{a} \frac{\dot{b}}{b} + \frac{\dot{a}}{a}
\frac{\dot{c}}{c} + \frac{\dot{b}}{b} \frac{\dot{c}}{c}  =
\Lambda + \kappa \left[\frac{1}{2}\Psi^2_0 (V + \dot{\phi}^2) + \frac{F^2_{12}}{2a^2 b^2} \left(1 -\lambda_1 \dot{\phi}^2 +\phi^2 \right) \right], \label{2Ein00}
\end{equation}
\begin{equation}
\frac{\ddot{b}}{b} \ + \frac{\ddot{c}}{c} \ + \frac{\dot{b}}{b}
\frac{\dot{c}}{c} \ =
\ \ \Lambda + \kappa \left[\frac{1}{2}\Psi^2_0 (V -\dot{\phi}^2) - \frac{F^2_{12}}{2a^2 b^2} \left( 1 + \lambda_1 \dot{\phi}^2  +  \phi^2 \right) \right],
\label{2Ein11}
\end{equation}
\begin{equation}
\frac{\ddot{a}}{a} \ + \frac{\ddot{c}}{c} \ + \frac{\dot{a}}{a}
\frac{\dot{c}}{c} \ =
 \ \ \Lambda + \kappa \left[\frac{1}{2}\Psi^2_0 (V -\dot{\phi}^2)  -\frac{F^2_{12}}{2a^2 b^2} \left(1+\lambda_1 \dot{\phi}^2+ \phi^2 \right)\right],
\label{2Ein22}
\end{equation}
\begin{equation}
\frac{\ddot{a}}{a} \ + \frac{\ddot{b}}{b} \ + \frac{\dot{a}}{a}
\frac{\dot{b}}{b} \ =
\ \ \Lambda + \kappa \left[\frac{1}{2}\Psi^2_0 (V -\dot{\phi}^2) +
\frac{F^2_{12}}{2a^2 b^2} \left( 1+\lambda_1 \dot{\phi}^2 + \phi^2 \right) \right]. \label{2Ein33}
\end{equation}
Clearly, the sources in the right-hand sides of (\ref{2Ein11}) and (\ref{2Ein22}) coincide, since both the original magnetic and the induced electric fields are directed along one axis ($0x^3$).
This is a good motivation to use the model with the local rotation symmetry (LRS), which assumes that $a(t){=}b(t)\neq c(t)$. Below we use this simplification and distinguish the longitudinal (along $0x^3$) and transversal (on the plane $x^10x^2$) dynamics; to describe them we introduce the longitudinal Hubble function $H_{||} {=} \frac{\dot{c}}{c}$, and the transversal Hubble function $H {=} \frac{\dot{a}}{a}{=}\frac{\dot{b}}{b}$.

\subsection{Constant solution with hidden $\lambda_1$ and $\lambda_2$}
\label{subsec32}

Let the potential of the axion field be quadratic in $\phi$, i.e., $V(\phi^2){=} m^2 \phi^2$.
Then for the special set of the guiding parameters of the model the total system of equations (\ref{cosm9})-(\ref{2Ein33}) admits the exact solution, for which the pseudoscalar field is constant $\phi(t){=}\phi_0 \neq 0$, and the scale factors $a$ and $b$ are constant, i.e., $a(t){=}a_0$, $b(t){=}b_0$, the only factor $c(t)$ being the function of time.
Indeed, the equation (\ref{cosm9}) is satisfied if
\begin{equation}
F^2_{12} = m^2 a^2_0 b^2_0 \Psi^2_0 \,.
\label{cosm310}
\end{equation}
In this case the equation (\ref{2Ein00}) determines the value $\phi_0$ by the relation
\begin{equation}
-\phi^2_0 = \frac{1}{2} + \frac{\Lambda}{\kappa m^2\Psi^2_0} <0    \,. \label{5Ein00}
\end{equation}
This is possible only for the anti-de Sitter-type model, when the cosmological constant is negative, and
\begin{equation}
 |\Lambda| > \frac{1}{2} \kappa m^2 \Psi^2_0 \,. \label{3Ein006}
\end{equation}
Because of relation (\ref{cosm310}) the equation (\ref{2Ein33}) turns into identity, and the equations (\ref{2Ein11}) and (\ref{2Ein22}) yield the key equation for the scale factor $c(t)$
\begin{equation}
\frac{\ddot{c}}{c} = - \Omega^2 \,, \quad \Omega^2 \equiv  |\Lambda| +  \frac{1}{2} \kappa m^2 \Psi^2_0  \,.
\label{3Ein11}
\end{equation}
Let us emphasize that the parameters $\lambda_1$ and $\lambda_2$ are hidden in this case: they do not appear in the equations since in all the key equations they are coefficients in front of the time derivative $\dot{\phi}$.
The solution to (\ref{3Ein11})
\begin{equation}
c(t)= c(0) \cos{\Omega t}+ \frac{\dot{c}(0)}{\Omega} \sin{\Omega t} \,,
\label{31Ein117}
\end{equation}
describes harmonic oscillations of the Universe in the direction along the magnetic and electric fields.

\subsection{Associated dynamic system}
\label{subsec33}

Let us consider the Bianchi-I model with LRS and introduce the following convenient dimensionless variables
\begin{equation}
x=\frac{a(t)}{a(t_0)}
\,, \quad
\frac{d}{dt} = xH \frac{d}{dx} \,, \label{3D1}
\end{equation}
\begin{equation}
X = \phi(x) \,, \quad Y= x \frac{d}{dx}\phi \,, \quad
Z=\frac{H^2}{H_0^2}
\,, \label{3D2}
\end{equation}
and auxiliary parameters
\begin{equation}
\nu^2  \equiv - \lambda_1 H_0^2 >0
\,, \quad
H_0 {=} H(t_0)
\,,
\label{3D3}
\end{equation}
\begin{equation}
\alpha \equiv \kappa \Psi_0^2
\,, \quad
\Lambda = \lambda H_0^2
\,, \quad
\beta \equiv  \frac{\kappa F^2_{12}}{H_0^2 a^4(t_0)}
\,.
\label{3D4}
\end{equation}
For the potential of the pseudoscalar field we use here the (simplest) linear ansatz
\begin{equation}
V(\phi^2)= H_0^2 \left (\gamma +  \mu^2 \phi^2 \right)\,,
\label{3D6}
\end{equation}
and formally put $\gamma {=}0$, since this constant can be included into an effective dimensionless cosmological constant $\lambda$.
In this terms the whole set of equations splits into one equation describing the longitudinal dynamics and
three-dimensional dynamic system for the description of the transversal dynamics.

\vspace{3mm}
\noindent
{\it Remark about initial data}

\noindent
The moment $t{=}t_0$ of cosmological time, for which $x{=}1$, relates to some
intermediate point, which splits the Universe history into two intervals: the second one, $1<x<\infty$, finishes by the so-called late-time period, the first interval,
$0<x<1$ starts with the so-called early Universe. Clearly, the replacement $x \to 1/\tilde{x}$ converts the last interval into $1<\tilde{x}<\infty$ and effectively changes
the time arrow. We choose the moment $t_0$ so that at $t{=}t_0$ the effective contribution of the magnetic field into the total energy of the Universe substratum is equal to the contribution of the dark matter. More precisely, we require that
\begin{equation}
m^2 \Psi_0^2 a^4(t_0) = F^2_{12}
\,,
\label{3D41}
\end{equation}
and thus $\beta {=} \alpha \mu^2$ (let us remind that $m^2{=}\mu^2 H_0^2$). The parameter $F_{12}$ related to the initial value of the magnetic field is not fixed, it is a fitting parameter.
In other words, the interval $0<x<1$ can be indicated as an epoch of magnetic field domination, while the interval $1<x<\infty$ is in these terms the epoch of dark matter domination. The initial data for other unknown functions, namely, $\phi(t_0)$, $\dot{\phi}(t_0)$, $a(t_0)$, $\dot{a}(t_0)$, $c(t_0)$, $\dot{c}(t_0)$, being necessary elements of the model, play different roles in the numerical and qualitative analysis. As we have mentioned above, the moment $t{=}t_0$ relates to $x{=}1$, thus the values
$\phi(t_0)$, $\dot{\phi}(t_0)$ are codified, in fact, in the parameters $X(1)$ and $Y(1)$; we listed these parameters in the captions for the Fig.1, Fig.3, Fig.5. The values $\dot{a}(t_0)$ and  $\dot{c}(t_0)$ are included into $H(t_0){=}\dot{a}(t_0)/a(t_0)$ and $H_{||}(t_0){=}\dot{c}(t_0)/c(t_0)$, respectively. The initial value of the transversal Hubble function $H(t_0)$ is not fixed, it is a fitting parameter. We analyzed numerically the ratio $H(t)/H(t_0)$, so that the initial value of this function is equal to one (see the finishing point of the curve displayed on the panel (a) of the Fig.4, and the starting point of the curve on the panel (a) of the Fig.3). In our computational scheme the initial value of the ratio $H_{||}(t_0)/H(t_0)$ (according to (\ref{2Ein00})) is not an arbitrary parameter, it is predetermined by the initial values $\phi(t_0)$, $\dot{\phi}(t_0)$ and by the values of the guiding parameters $\alpha$, $\nu$, $\mu$ (see the right-hand side of (\ref{2Ein00})). Mentioned parameters are listed in the captions to the Fig.1, Fig.3 and Fig.5; the corresponding initial value for the ratio $H_{||}(t_0)/H(t_0)$ can be found from the plots. Finally, the initial values of the scale factors, $a(t_0)$ and $c(t_0)$, are hidden parameters in our computational scheme, they also can be considered as fitting parameters.


\subsubsection{Evolution of the longitudinal scale factor}
\label{subsubsec331}

The longitudinal Hubble function $H_{||}{=}\frac{\dot{c}}{c}$ can be found from the equation (\ref{2Ein00}) (with $a{=}b$).  In terms of variable $x$ the equation for $c(x)$ reads
\begin{equation}
x \frac{d}{dx} \log{c} = \frac{V_{c}}{2Z}\,,
\label{03D5}
\end{equation}
where $V_{c}(x)$ is given by the function
\begin{equation}
V_{c} {=} \lambda {-} Z {+} \frac{\alpha}{2}\left[\mu^2 X^2 {+} Z Y^2 {+} \frac{\mu^2}{x^4} \left(1 {+} \nu^2 Z Y^2 {+} X^2 \right)\right].
\label{3D5}
\end{equation}
The dimensionless function, which we are interested to find, is
$\frac{H_{||}}{H_0}{=}\frac{ V_{c}}{2\sqrt{Z}} $.

\subsubsection{Transversal 3D-dynamic system}
\label{subsubsec332}

The dimensionless functions $X$, $Y$ and $Z$, introduced by (\ref{3D2}), are coupled by the set of equations
\begin{equation}
x\frac{d}{dx} X = Y
\,, \quad
x \frac{d}{dx} Y = V_{Y}
\,, \quad
x \frac{d}{dx} Z = V_{Z} + \lambda \,,
\label{3D7}
\end{equation}
where $V_{Y}$ and $V_{Z}$ are the following nonlinear functions of variables $x$, $X$, $Y$, $Z$, and of the guiding parameters
$\nu$, $\mu$, $\alpha$, $\lambda$:
\begin{equation}
V_{Y}{=} \frac{X \mu^2 (1{-} x^4)}{Z(x^4 {+} \nu^2 \mu^2)}
{-} Y \left[\frac{V_{c} {+}V_{Z}}{2Z} {+} 2 \frac{(x^4{-} \nu^2 \mu^2)}{(x^4 {+}\nu^2 \mu^2)}\right],
\label{3D8}
\end{equation}
\begin{equation}
V_{Z} {=} \frac{\alpha}{2}\left(\mu^2 X^2 {-} ZY^2 \right) {-}3 Z
{+}\frac{\alpha \mu^2}{2x^4} \left(1{-}\nu^2 Z Y^2 {+} X^2 \right).
\label{3D11}
\end{equation}
Below we analyze this dynamic system qualitatively and numerically.

\subsubsection{Numerical analysis: \\ general properties of magnetic, electric and pseudoscalar fields}
\label{subsubsec333}

Numerical analysis of the model is fulfilled for various values of the dimensionless guiding parameters $\nu, \mu, \alpha, \lambda$ and initial values $X(1), Y(1)$. Below we discuss main tendencies and typical results using the most illustrative examples. Let us mention that the magnetic field $B(x){=}\frac{B(1)}{x^2}$ (see (\ref{cosm555})) decreases monotonically, and its profile does not require illustrations. The electric field, induced by the axion - photon interactions, $E(x)=B(x) \phi(x)$, (see (\ref{cosm556})), in fact inherits the properties of the axion field $\phi(x)$, and we do not display the plot of its profile.
Typical behavior of the pseudoscalar field $\phi(x)$ (obtained from the solution for $X(x)$) is displayed on the Fig.1 for the whole interval $0<x<\infty$.
The line number 2 presents the plot $\phi(x)$ for the model with $\lambda_1{=}\lambda_2{=}0$, i.e., for the case when the gradient-type terms in the Lagrangian (\ref{action}) are absent.
Clearly, late-time evolution of the pseudoscalar field does not feel the presence or absence of the terms with $\lambda_1$ and $\lambda_2$; the interactions described by these terms are important in the early Universe. It is important to stress that when $\lambda_1{=}\lambda_2{=}0$, the modulus $|\phi|$ is infinite at $x \to 0$ (see the second line), while for gradient-type models the pseudoscalar field typically tends to finite value at $x \to 0$. Another typical feature is a non-monotonic character of the $\phi(x)$ profile: the pseudoscalar (axion) field  grows in the early Universe, reaches the maximum value $\phi_{({\rm max})}(\nu,\mu,\alpha,\lambda)$ (at $x>1$), then decreases at $x \to \infty$ in a quasi-oscillatory regime.

\begin{figure}
[h]
\includegraphics[width=0.51\textwidth]{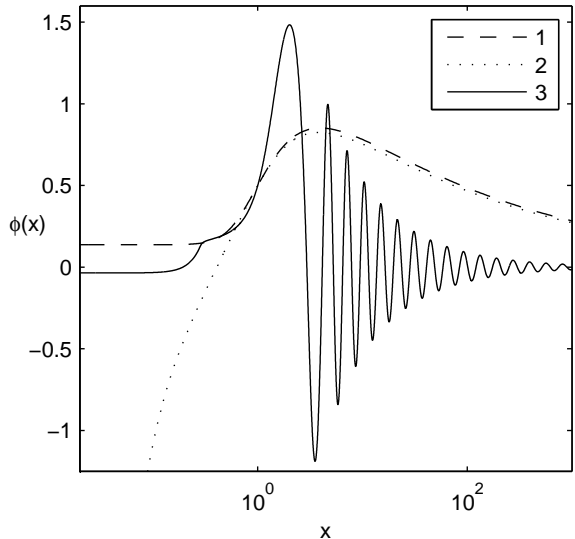}
\caption{The plots of the pseudoscalar field, $\phi(x)$. The first line relates to $\alpha{=}0.1$, $\nu^2{=}1$, $\lambda{=}1$, $\mu^2{=}0.1$, $X(1){=}0.5$, $Y(1){=}0.5$. The example 2 differs from the example 1 by the condition $\lambda_1{=}0{=}\lambda_2$, i.e., $\nu^2{=}0$.  The line 3, with the guiding parameters $\alpha{=}0.01$, $\nu^2{=}3$, $\lambda{=}1$, $\mu^2{=}100$, $X(1){=}0.5$, $Y(1){=}1$, demonstrates a quasi-oscillatory regime of the asymptotic behavior at $x \to \infty$.
}
\label{fig:1}
\end{figure}

Concerning the plots for the longitudinal $H_{||}{=}\frac{\dot{c}}{c}$ and transversal $H{=}\frac{\dot{a}}{a}$ Hubble functions, below we consider them separately for the intervals $0<x<1$ and $1<x<\infty$, since the appropriate visualization of the profile details requires to use different scales for these plots.

\subsection{The epoch of dark matter domination ($x>1$): \\ Asymptotic behavior and late-time isotropization}
\label{subsec34}

The terms containing $F^2_{12}a^{{-}2}b^{{-}2}$ in the key equations of the model become vanishing at
$a \to \infty$, $b(t) \to \infty$ (see, e.g., (\ref{cosm9})-(\ref{2Ein33})). Both remaining sources of the gravity field, the axionic dark matter and dark energy ($\Lambda$ term) are spatially isotropic. Comparing the left-hand sides of the equations (\ref{2Ein11})- (\ref{2Ein33}), and using the LRS ansatz, one can show that the appropriate solution in this limit is the model with
$\frac{\dot{a}}{a}{=}\frac{\dot{b}}{b}{=}\frac{\dot{c}}{c}$. In other words, the scale factors $a{=}b$ and $c$ become proportional to one another, thus, with rescaling of the coordinate $x^3$, we can obtain the asymptotically isotropic model with $a(t){=}b(t){=}c(t) \to \infty$. In this limit we deal with standard isotropic homogeneous FLRW-type model, in which the cosmological constant and the axionic dark matter predetermine the Universe evolution.

\subsubsection{Solutions with de Sitter asymptote}
\label{subsubsec341}

The late-time isotropization at $\Lambda >0$ can be studied using the following qualitative arguments. At $x\to \infty$ the associated dynamic system becomes autonomous
\begin{equation}
\frac{d}{d\tau} X = Y
\,, \quad \tau \equiv \log{x}\,,
\label{3D01}
\end{equation}
\begin{equation}
\frac{d}{d\tau} Y = {-}\mu^2 \frac{X}{Z} {-} \frac{Y}{2Z} \left(\lambda {+} \alpha \mu^2 X^2 \right) \,,
\label{3D02}
\end{equation}
\begin{equation}
\frac{d}{d\tau}Z = \lambda + \frac{\alpha}{2}\left(\mu^2 X^2 {-} ZY^2 \right) {-}3 Z \,,
\label{3D03}
\end{equation}
\begin{equation}
\frac{d}{d\tau} \log{c} =
\frac{1}{2Z} \left[\lambda - Z + \frac{\alpha}{2}\left(\mu^2 X^2 + Z Y^2 \right) \right]
\,.
\label{3D04}
\end{equation}
The autonomous subsystem (\ref{3D01})-(\ref{3D03}) has only one stationary point
\begin{equation}
X_{*} =0 \,, \quad Y_{*} = 0 \,,  \quad Z_{*} = \frac{\lambda}{3}
\,.
\label{3D05}
\end{equation}
In the vicinity of this point the equation (\ref{3D03}) yields
\begin{equation}
Z(x) \to  \frac{\lambda}{3} + C_3 \ x^{-3}
\,, \quad H(x) \to \sqrt{\frac{\Lambda}{3}} + \tilde{C}_3 \ x^{-3} \,,
\label{3D06}
\end{equation}
and we obtain the de Sitter asymptote
\begin{equation}
\tau = \sqrt{\frac{\Lambda}{3}} \ t \,, \quad a(t) \to a(t_0)\exp{\left[\sqrt{\frac{\Lambda}{3}}(t-t_0)\right]} \,.
\label{3D07}
\end{equation}
The phase plane $X0Y$ describes (in the appropriate dimensionless variables) the behavior of $\phi$ and $\dot{\phi}$. It is easy to show that the
stationary point $(0,0)$ on this phase plane is a stable node, when $\Lambda \geq \frac{16}{3}m^2$, or a stable focus, when $\Lambda<\frac{16}{3}m^2$. In both cases
the pseudoscalar field itself and its time derivative vanish exponentially with time. Fig.2 displays the example of a phase portrait of this 2D dynamic system.

\begin{figure}
[h]
\includegraphics[width=0.49\textwidth]{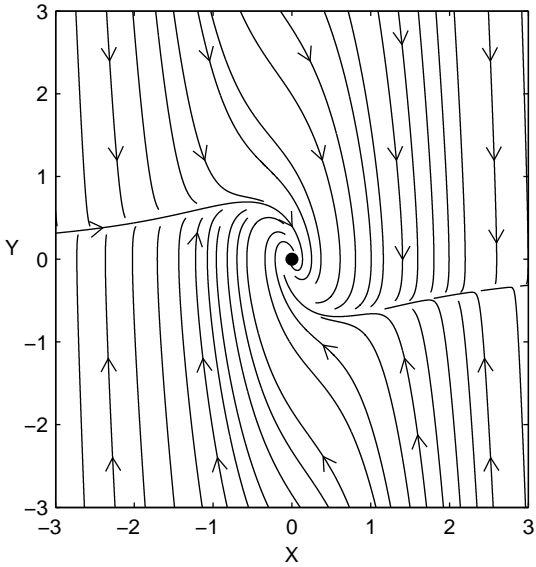}
\caption{Phase portrait of 2D dynamic system at $x \to \infty$ reconstructed for $\alpha{=}2$, $\nu{=}1$, $\lambda{=}1$, $\mu{=}1$. The stable stationary point corresponds to $X{=}0$ and $Y{=}0$, i.e., $\phi(\infty){=}0$ and $\dot{\phi}(\infty){=}0$.}
\label{fig:2}
\end{figure}

Concerning the longitudinal dynamics, in the vicinity of the stationary point the
scale factor $c(t)$ behaves as follows
\begin{equation}
\frac{d}{d\tau} \log{c} \to 1 \,, \quad c(t) \to c(t_0)\exp{\left[\sqrt{\frac{\Lambda}{3}}(t-t_0)\right]} \,,
\label{3D08}
\end{equation}
displaying explicitly the isotropization law.

\subsubsection{Special solution with anti-de Sitter asymptote}
\label{subsubsec342}

When $\Lambda$ is negative the special solution of the isotropic model exists, for which the pseudoscalar potential is linear in time. Using the simplest potential
$V(\phi^2){=} m^2 \phi^2$ we can reduce the master equations for the isotropic stage  to the pair of equations containing, first, the standard equation
\begin{equation}
\ddot{\phi} + 3 H \dot{\phi} + m^2 \phi = 0
\label{7cosm101}
\end{equation}
for the pseudoscalar (axion) field, second, the equation
\begin{equation}
3 H^2 =
\Lambda +  \frac{1}{2}\kappa \Psi^2_0 (m^2 \phi^2+\dot{\phi}^2)
\label{7Ein001}
\end{equation}
for the Hubble function $H(t)\equiv \frac{\dot{a}}{a}$. We would like here to attract the attention to the specific exact solutions linear in time
\begin{equation}
\phi(t) = \phi(0) \left(1-\frac{t}{t_{*}} \right) \,, \quad t_{*} \equiv \sqrt{\frac{3 \kappa}{2m^2}}\Psi_0 \phi(0) \,,
\label{7cosm201}
\end{equation}
\begin{equation}
H(t) = \sqrt{\frac{m^2 \kappa}{6}}\Psi_0 \phi(0) \left(1-\frac{t}{t_{*}} \right) \,,
\label{7cosm2016}
\end{equation}
which hold, if $\Lambda{=}{-}\frac{1}{3}m^2$.
At the moment $t=t_{*}$ the pseudoscalar field and the Hubble function change their signs.
This solution gives the Gaussian function for the scale factor
\begin{equation}
a(t)= a(t_{*}) \exp\left\{-\frac{1}{6}m^2 (t-t_{*})^2 \right\} \,.
\label{7cosm203}
\end{equation}
For the time interval $t<t_{*}$ this function grows, and the acceleration parameter
\begin{equation}
-q(t) \equiv \frac{\ddot{a}}{a H^2}= 1+ \frac{\dot{H}}{H^2}  = 1 - \frac{3}{m^2 (t-t_{*})^2}
\label{7cosm204}
\end{equation}
is positive, when
$t<t_{*}- \sqrt{3}/m $.
Clearly, according to this model the accelerated expansion of the Universe will be changed by decelerated expansion, then
the scale factor will reach the maximal value $a_{({\rm max})}=a(t_{*})$,
and finally, the epoch of collapse will take place.

\subsubsection{Numerical analysis of the transversal and longitudinal Hubble functions at $x>1$}
\label{subsubsec343}

The results of numerical analysis of the dynamic system (\ref{03D5})-(\ref{3D11}) for the interval $1<x<\infty$ are presented in the Fig.3. The panel (a)
illustrates the behavior of the ratio $H/H_0$, obtained as square root of the function $Z$, where the transversal Hubble function is defined as $H{=}\frac{\dot{a}}{a}$.
Similarly, the panel (b) displays the behavior of the longitudinal Hubble function $H_{||}{=}\frac{\dot{c}}{c}$. Both graphs illustrate the tendency to isotropization, i.e., both $H$ and $H_{||}$ tend asymptotically to the value $H_{\infty} {=} \sqrt{\Lambda/3}$. The type of graph is predetermined by the value of $\lambda$: when $\lambda>3$ the plots start below the asymptote, when $\lambda<3$, they start above this asymptote; we used the second case for the illustration. Depending on the guiding parameters $\nu, \mu, \alpha, \lambda$ the profiles $H(x)$ and $H_{||}(x)$ can have zero, one, two, etc. extrema. The lines indicated by the number 3 relate to the model with $\lambda_1{=}0{=}\lambda_2$.

\begin{figure}
[t]
\includegraphics[width=0.49\textwidth]{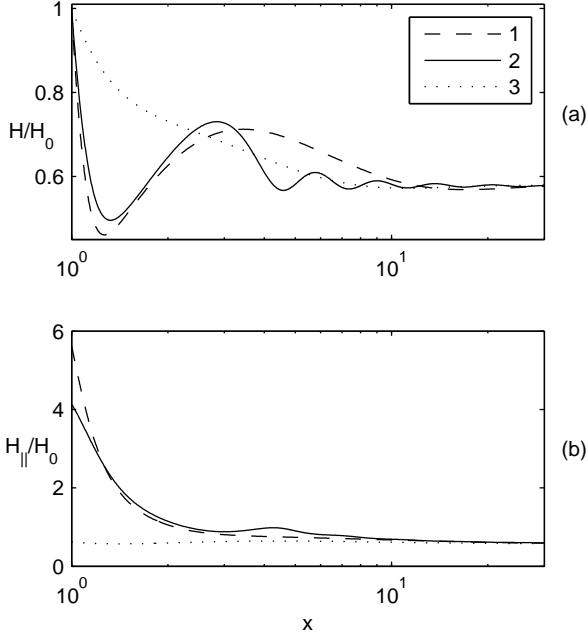}
\caption{Behavior of the Hubble functions in the epoch of dark matter domination ($1<x<\infty$). Panels (a) and (b) display the functions $H/H_0$ and $H_{||}/H_0$, respectively. The lines indicated by the number 1 relate to
$\alpha{=}1$, $\nu^2{=}20$, $\lambda{=}1$, $\mu^2{=}1$, $X(1){=}0.5$, $Y(1){=}1$; for the lines 2 the parameters $\alpha{=}0.05$, $\nu^2{=}15$, $\lambda{=}1$, $\mu^2{=}20$, $X(1){=}0.5$, $Y(1){=}1$ are used; the line 3 illustrates the model with $\lambda_1{=}0{=}\lambda_2$.}
\label{fig:3}
\end{figure}

\subsection{The epoch of magnetic field domination ($x<1$): \\ Anisotropic early Universe}
\label{subsec35}

When the values of the scale factors are small, the terms with $F^2_{12}$ dominate over the terms with $\Psi^2_0$ in the key equations
(\ref{cosm9})-(\ref{2Ein33}). We consider the dynamic system (\ref{03D5})-(\ref{3D11})
numerically in the interval $0<x<1$; the results are illustrated by the Fig.4, Fig.5 and Fig.1.

\begin{figure}
[h]
\includegraphics[width=0.49\textwidth]{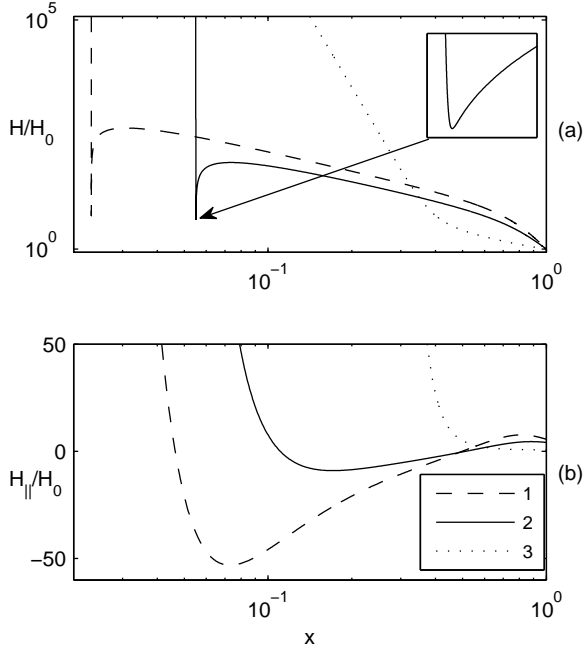}
\caption{Behavior of the Hubble functions in the epoch of magnetic field domination ($0<x<1$). The guiding parameters are the same as on the Fig.3. Additional window visualizes the behavior of the curve $H(x)/H_0$ near the local minimum.}
\label{fig:4}
\end{figure}

The distinguished feature of the graphs is the following: for various sets of guiding parameters there exists a pair of extremum points on the curve $H(x)/H_{0}$, the minimum (at $x{=}x_{({\min})}$) and the maximum (at $x{=}x_{({\max})}$). Changing the set of guiding parameters one can shift these extrema (see, e.g., first and second lines on the Fig.4). In the interval $0<x<x_{({\min})}$ the function $H(x)/H_{0}$ grows infinitely at $x \to 0$. In the interval $x_{({\min})}<x<x_{({\max})}$ the function $H^2(x)$ can be effectively fitted by the function $H^2 \propto x^{{-}4}$, while in the interval $x_{({\max})}<x<1$ it behaves as $H^2 \propto x^{{-}3}$. The reconstruction of the vector field for the integral curves on the phase plane $Z0Y$ at small $x$ (see Fig.5) confirms, that, indeed, the quantity $Z{=} H^2(x)/H^2_0$ decreases, then passes through the minimum and, finally, tends to infinity at $x \to 0$. Auxiliary dashed lines on the Fig.5 mark the positions of the minimum, clearly, the minimum shifts to smaller values of $x$ when $Y{=}x \frac{d \phi}{dx}$ decreases.
Such behavior can be explained qualitatively as follows. At $x \to 0$ the dynamic system (\ref{3D5})-(\ref{3D11}) can be transformed into
\begin{equation}
x\frac{d}{dx} X = Y
\,, \label{3D97}
\end{equation}
\begin{equation}
x \frac{d}{dx} Y = Y \left[4 - \frac{\alpha \mu^2}{2x^4Z} \left(1 {+} X^2 \right) \right] + \frac{X}{\nu^2 Z }
\,,
\label{3D87}
\end{equation}
\begin{equation}
x \frac{d}{dx} Z = - Z \left[3 + \frac{\alpha \nu^2 \mu^2}{2x^4} Y^2 \right]+ \frac{\alpha \mu^2}{2x^4} \left(1{+} X^2 \right)
 \,,
\label{3D77}
\end{equation}
and the approximate solutions have the form
\begin{equation}
X  \to X_0 \left[1{+} \frac{x^4}{2\alpha \nu^2 \mu^2 (1{+}X^2_0)} \right] \,,  \quad
Y  \to Y_0  x^5 + \frac{2X_0}{\alpha \nu^2 \mu^2 (1{+}X^2_0)} \ x^4
\,, \label{XYZ11}
\end{equation}
\begin{equation}
Z \to  \left[1{+} \frac{1}{2}\alpha \mu^2 (1{+}X^2_0) \right]x^{{-}3} {-} \frac{1}{2}\alpha \mu^2 (1{+}X^2_0) x^{{-}4}
\,. \label{XYZ2}
\end{equation}
Here $X_0{=}X(0)$, and we assume that the fitted integral curve passes through the point $Z(1){=}1$.
The local maximum of this curve corresponds to the parameters
\begin{equation}
x_{({\rm max})} {=} \frac{2\alpha \mu^2 (1{+}X^2_0)}{3\left[1{+}\frac{1}{2}\alpha \mu^2 (1{+}X^2_0)\right]}
\,,  \quad Z_{({\rm max})} {=} \frac{\alpha \mu^2 (1{+}X^2_0)}{6x^4_{({\rm max})}}
\,. \label{XYZ4}
\end{equation}
The formulas (\ref{XYZ11})-(\ref{XYZ4}) give a nice qualitative illustration of the results obtained numerically.

\begin{figure}
[h]
\includegraphics[width=0.49\textwidth]{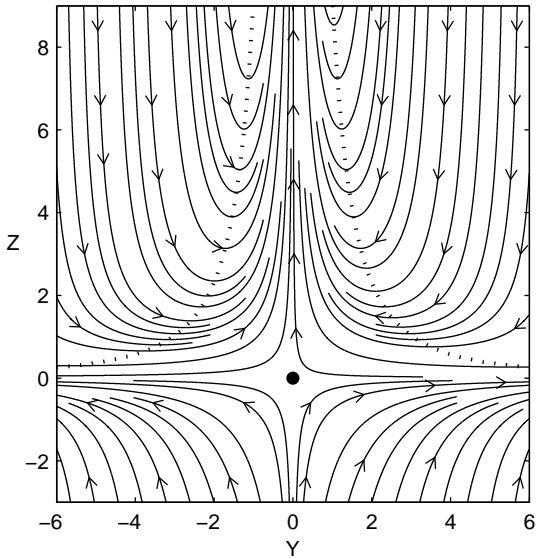}
\caption{Local fragment of the vector field picture for the integral curves on the phase plane $Z0Y$ at $x \to 0$. The point $(0,0)$ plays a role of a saddle point. Dashed lines indicate the points, which relate to the minimum positions of the curves depicted on the Fig.4. The illustrative parameters are chosen as follows
$\alpha{=}15$, $\nu^2{=}1$, $\lambda{=}1$, $\mu^2{=}0.01$, $X(1){=}0.5$, $Y(1){=}{-2}$.
}
\label{fig:5}
\end{figure}

\section{Discussion}
\label{secDISS}

Extended Einstein-Maxwell-axion model established in the work inherits main properties of the standard covariant model of pseudoscalar - photon coupling and introduces new interesting details; it is  worth studying these new features in the cosmological context, taking into account (hypothetic) axionic nature of the dark matter. The analysis of the first subset of the total self-consistent system of master equations, namely, of the equations of extended axion electrodynamics (\ref{eld1}),(\ref{eld2}) and (\ref{Emaxstar}), shows that in addition to the well-known magneto - electric effect the birefringence phenomenon is admissible. This phenomenon can be visible, when the gradient four-vector of the pseudoscalar field, $\nabla_k \phi$ has non-vanishing spatial part (see, e.g., the formulas (\ref{eld11}) and (\ref{eld12}) for the anisotropic permittivity tensors). In the framework of spatially homogeneous Bianchi-I cosmological model the birefringence does not appear, however, the magnetic field produces the electric field colli\-near to the magnetic one, due to the axion-photon interactions. This axionically induced electric field (see (\ref{cosm301})) can be an interesting addition to the large-scale cosmological magnetic field, the cosmic substrate, which seems to be very important for the Universe structure and evolution (see, e.g., \cite{TurnerR,Od1}).

The extended equation of the pseudoscalar field evolution (see (\ref{ps1}) with (\ref{ps2})) differs from the standard equation by the propagation operator; this means, in particular, that the velocity of pseudoscalar waves propagation depends on the direction and intensity of electric and magnetic fields. In the cosmological context the most important consequence of the model extension for the axion field evolution is the following. The pseudoscalar source in the right-hand side of (\ref{ps1}) being proportional to the scalar product of the electric and magnetic field vectors, is equal to zero, when the cosmological model includes the magnetic field only. Since due to the axion-photon interactions the initial magnetic field produces the electric field, this pseudoscalar source happens to be switched on in the early Universe and provides the (essential) growth of the axion field in the epoch of magnetic field domination (see, e.g., the Fig.1). One can expect that this (magneto-electric) mechanism of the pseudoscalar field production plays a nontrivial role in the process of the background (relic) axions creation, which is important for understanding of the dark matter problem.

Extended gravity field equations (\ref{gr1})-(\ref{gr5}) contain new source-terms, which modify the gravity field dynamics in the presence of the axion-photon interactions of the gradient type.
In the cosmological context such modifications are shown to be essential, when the contribution from the magnetic field into the total cosmic stress-energy tensor dominates over the contribution of the dark matter, i.e., in the early Universe (see Fig.4).

The gradient model of the axion-photon interaction is the two-parameter one, i.e., two coupling constants $\lambda_1$ and $\lambda_2$ have been phenomenologically introduced, when we formulated generic Lagrangian with gradient-type cross-invariants (\ref{action}). Keeping in mind the principle, that it is preferable not to introduce ``New Constants of Nature'', we propose to use the following relationships. First of all, in order to guarantee that the electric field, produced by interaction of the initial magnetic field and axions, is nonsingular at least in two physically different situations (cosmological and static spherically symmetric models), we have shown that $\lambda_1$ and $\lambda_2$ should satisfy the linear relation $\lambda_1 {+}\frac{1}{2}\lambda_2{=}0$ (see (\ref{cosm301}) and (\ref{sss4})); thus the two-parameter model was reduced to the one-parameter. Similarly, the axion field is guaranteed to be nonsingular (see (\ref{cosm9})), when $\lambda_1$ is negative, i.e., $\lambda_1{=}-\nu^2$.
There are few possibilities to express the parameter $\nu^2$ in terms of well-known physical constants. For instance, keeping in mind (\ref{3D8}) one can put $\nu^2{=}1/\mu^2$. Since in this case the coupling constant is reciprocal to the axion mass in square, we have a direct analogy with the Drummond-Hathrell model of the nonminimal coupling of gravitational and electromagnetic fields \cite{DH} (let us remind that the mentioned coupling constant is reciprocal to the electron mass in square, or more precisely, to the square of the Compton radius).

Another possibility to choose the constant $\nu^2$ by a physically motivated manner follows from the analysis of the Fig.4(a): we can choose the parameter $\nu^2$
so that the moment $t_{({\rm min})}$, at which the curve $H(x)$ reaches the local minimum, will coincide with the Planck time $t_{({\rm Planck})}$.
In this sense, the fragment of the curve $H/H_0$ between the local minimum and maximum, displayed on the Fig.4(a), can be interpreted in terms of inflationary - type evolution of the anisotropic
early Universe. According to (\ref{XYZ4}) the maximum value of the transversal Hubble function $H_{({\rm max})}{=} H_0 \sqrt{Z_{({\rm max})}}$ depends on the choice of the guiding
parameters $\alpha \nu^2$, $X_0$ and $H_0$. For the parameters listed in the caption the dashed curve has the maximum estimated by the value
$H_{({\rm max})} \simeq 10^3 H_0$; since $H_{({\rm max})} \propto x^{{-2}}_{({\rm max})}$ (see (\ref{XYZ4})), this value can be made as large as necessary,
by using small values of $x_{({\rm max})}$. It is important to emphasize, that the reference curve related to the model with $\lambda_1{=}\lambda_2{=}0$ has no extrema and thus
can not be used for modeling of the inflationary-type evolution. Clearly, in order to use the WMAP data  \cite{7Y} for the fitting of the parameters of our model,
we should estimate the reference value $H_{({\rm isotrop})}$ at the moment $t_{({\rm isotrop})}$, which relates to the starting point of the isotropic expansion of the Universe.
According to the Fig.3, for the given guiding parameters the isotropic epoch with $H \simeq H_{||}$ starts at $x \simeq 10$. We hope to present the results of the fitting procedure in the next work.

Returning to the questions posed in the beginning of Section \ref{sec3}, one can conclude the following. First, indeed,  the axion-photon
coupling in the presence of a strong cosmological magnetic field can produce a substantial growth of the pseudoscalar (axion) field $\phi$ (see Fig.1, the fragments of the curves for $x<1$),
thus providing a sensible number of cold axions in our (late-time) epoch. Second, the large number of axions produced in the epoch of the magnetic field domination can accelerate the process
of the Universe isotropization (see Fig.3), and this isotropization process is multi-stage.

We expect that the study of the gradient model of the axion-photon coupling will be interesting also in the context of astrophysical applications, and intend to consider static configurations of axionically coupled electric and magnetic fields in a future paper.

\vspace{5mm}

\noindent
{\bf Acknowledgements}

\noindent
This work was supported by the Federal Targeted Programme ``Scientific and Scientific - Pedagogical Personnel of the Innovative Russia''
(grants Nos 16.740.11. 0185 and  14.740.11. 0407), and by the Russian Foundation for Basic Research (grants Nos. 11-02-01162 and 11-05-97518 - p-center-a).
AB is grateful to Professor Wei-Tou Ni for stimulating discussions and helpful advices.


\begin{thebibliography}{}

\bibitem{Peccei0} R.D. Peccei, H.R. Quinn, {\it CP conservation in the presence of pseudoparticles},
Phys. Rev. Lett. {\bf 38}, 1440 (1977)

\bibitem{Weinberg0} S. Weinberg, {\it  A new light boson?},  Phys. Rev. Lett. {\bf 40}, 223 (1978)

\bibitem{Wilczek0}  F. Wilczek, {\it Problem of strong P and T invariance in the presence of instantons}, Phys. Rev. Lett. {\bf 40}, 279 (1978)

\bibitem{Turner} M.S. Turner,  {\it Windows on the axion}, Phys. Rep. {\bf 197}, 67 (1990)

\bibitem{Raffelt} G.G. Raffelt, {\it Astrophysical methods to constrain axions and other novel particle phenomena}, Phys. Rep. {\bf 198}, 1 (1990)

\bibitem{Kallosh} R. Kallosh, A. Linde, D. Linde, L. Susskind,  {\it Gravity and global symmetries }, Phys. Rev. {\bf D 52}, 912 (1995)

\bibitem{Peccei2} R.D. Peccei, {\it The Strong CP problem and axions }, Lect. Notes
Phys. {\bf 741}, 3 (2008)

\bibitem{Battesti} R. Battesti et al, {\it Axion searches in the past, at present, and in the near future}, Lect. Notes Phys. {\bf 741}, 199 (2008)

\bibitem{Ni0I} W.-T. Ni, {\it Equivalence principles and electromagnetism}, Phys. Rev. Lett. {\bf 38}, 301 (1977)

\bibitem{Wilczek} F. Wilczek,  {\it Two applications of axion electrodynamics},
 Phys. Rev. Lett. {\bf 58}, 1799 (1987)

\bibitem{rot2} S.M. Carroll , G.B. Field, R. Jackiw, {\it Limits on a Lorentz- and parity-violating
modification of electrodynamics}, Phys. Rev. {\bf D 41}, 1231 (1990)

\bibitem{HO2} F.W. Hehl, Yu.N. Obukhov, {\it Measuring a piecewise
constant axion field in classical electrodynamics}, Phys.
Lett. {\bf A 341}, 357 (2005)

\bibitem{Itin} Y. Itin, {\it Wave propagation in axion electrodynamics},
 Gen. Relat. Grav. {\bf 40}, 1219 (2008)

\bibitem{19} W.-T. Ni,  {\it From equivalence principles to cosmology: cosmic polarization rotation, CMB observation,
neutrino number asymmetry, Lorentz invariance and CPT}, Prog.
Theor. Phys. Suppl. {\bf 172}, 49 (2008)

\bibitem{Ni11} W.-T. Ni,  {\it Foundations of Electromagnetism, Equivalence Principles and Cosmic Interactions}, (2011) arXiv:1109.5501

\bibitem{Claus1} R.A. Puntigam, C. L\"ammerzahl, F.W. Hehl,  {\it Maxwell's theory on a post-Riemannian spacetime and the equivalence principle},
Class. Quantum Grav. {\bf 14}, 1347 (1997)


\bibitem{Od0} S. Nojiri, S.D. Odintsov, S. Ogushi, A. Sugamoto, M. Yamamoto,
{\it Axion-dilatonic conformal anomaly from Ads/CFT correspondence},
 Phys. Lett. {\bf B 465}, 128 (1999)

\bibitem{EMA4} M. Halilsoy, I. Sakalli, {\it Collision of electromagnetic shock waves coupled with
axion waves: An Example}, Class. Quantum Grav. {\bf 20}, 1417 (2003)


\bibitem{EMA3} B.A. Bassett, M. Kunz, {\it Cosmic acceleration vs axion-photon mixing},
 Astrophys. J. {\bf 607}, 661 (2004)

\bibitem{Claus2} C. L\"ammerzahl, A. Macias, H. M\"uller,  {\it Lorentz invariance violation and charge (non-)conservation: A general theoretical
frame for extensions of the Maxwell equations},
Phys. Rev. {\bf D 71}, 025007 (2005)

\bibitem{EMA5} L.A. L\'{o}pez, N. Bret\'{o}n, {\it Asymptotic singular behaviour of
inhomogeneous cosmologies in Einstein-Maxwell-dilaton-axion theory},
Gen. Relat. Grav. {\bf 39}, 153 (2007)

\bibitem{HO2008} F.W. Hehl, Yu.N. Obukhov, {\it Equivalence principle and electromagnetic field: no birefringence, no dilaton, and no axion}, Gen. Rel. Grav. {\bf 40}, 1239 (2008)

\bibitem{Matos} T. Matos, G. Miranda, R. Sanchez-Sanchez, P. Wiederhold, {\it Class of Einstein-Maxwell-Dilaton-Axion Space-Times}, Phys. Rev. {\bf D 79}, 124016 (2009)

\bibitem{Zavattini1} E. Zavattini et al (PVLAS Collaboration), {\it Experimental observation of optical rotation generated in vacuum by a magnetic field},  Phys. Rev. Lett.
{\bf 96}, 110406 (2006)

\bibitem{Ni22} S.-J. Chen, H.-H. Mei, W.-T. Ni, {\it Q \& A experiment to search for vacuum dichroism,
pseudoscalar - photon interaction and millicharged fermions},
Mod. Phys. Lett. {\bf A 22}, 2815 (2007)

\bibitem {Zavattini2} E. Zavattini et al, {\it New PVLAS results and limits on magnetically induced optical
rotation and ellipticity in vacuum}, Phys. Rev. {\bf D 77},
032006 (2008)

\bibitem{Battesti24} R. Battesti et al, {\it The BMV experiment: a novel apparatus to study the propagation
of light in a transverse magnetic field},  Eur. Phys. J.  {\bf D 46}, 323 (2008)

\bibitem{Ni25} W.-T. Ni, {\it Cosmic polarization rotation, cosmological models, and the detectability
of primordial gravitational waves}, Int. J. Mod. Phys. {\bf A
18} \& {\bf 19}, 3493 (2009)

\bibitem{Shellard} E.P.S. Shellard, R.A. Battye, {\it On the origin of dark matter axions},
Phys. Rept. {\bf 307}, 227 (1998)

\bibitem{Duffy} L.D. Duffy, K. van Bibber, {\it Axions as Dark Matter Particles}, New J. Phys. {\bf 11}, 105008 (2009)

\bibitem{Sikivie} P. Sikivie, Q. Yang, {\it Bose-Einstein Condensation of Dark Matter Axions},  Phys. Rev. Lett. {\bf 103}, 111301 (2009)

\bibitem{Visinelli} L. Visinelli, P. Gondolo, {\it Axion cold dark matter in non-standard cosmologies}, Phys. Rev. {\bf D 81}, 063508 (2010)

\bibitem{BaWTNi}  A.B. Balakin, W.-T. Ni, {\it Non-minimal coupling of photons and axions}, Class. Quantum Grav. {\bf 27}, 055003 (2010)

\bibitem{BaWTNi2} W.-T. Ni, A.B. Balakin, H.-H. Mei, {\it Pseudoscalar-photon interactions, axions, non-minimal extensions, and their empirical constraints from observations}, in {\it
Proceedings of the Conference in Honour of Murray Gell-Mann's 80th birthday: Quantum mechanics, Elementary Particles, Quantum Cosmology and Complexity}, Singapore, 2010,
(World Scientific Publishing Co., Singapore, 2011), p. 526

\bibitem{EM}  A.C. Eringen, G.A. Maugin,
{\it Electrodynamics of continua} (Springer-Verlag, New York, 1989)

\bibitem{LLP}  L.D. Landau, E.M. Lifshitz, L.P. Pitaevskii,
{\it Electrodynamics of continuous media} (Butterworth
Heinemann, Oxford, 1996)

\bibitem{HehlObukhov} F.W. Hehl, Yu.N. Obukhov,
{\it Foundations of classical electrodynamics: Charge, flux, and
metric} (Birkh\"auser, Boston, 2003).

\bibitem{Ba07} A.B. Balakin, {\it Magnetic relaxation in the Bianchi-I Universe}, Class. Quantum Grav. {\bf 24}, 5221 (2007)

\bibitem{Ba071} A.B. Balakin, {\it Extended Einstein-Maxwell model}, Grav.Cosmol {\bf 13}, 163 (2007)

\bibitem{derc1} L. Amendola, {\it Cosmology with non-minimal derivative couplings},
 Phys. Lett. {\bf B 301}, 175 (1993)

\bibitem{derc2} S. Capozziello, G. Lambiase, {\it  Non-minimal derivative coupling and the recovering of cosmological constant}, Gen. Relat. Grav. {\bf 31}, 1005 (1999)

\bibitem{derc3} S. Capozziello, G. Lambiase, H.-J. Schmidt, {\it  Non-minimal derivative couplings
and inflation in generalized theories of gravity}, Annalen
Phys. {\bf 9}, 39 (2000)

\bibitem{BZD1} A.B. Balakin, H. Dehnen, A.E. Zayats, {\it Non-minimal Einstein-Yang-Mills-Higgs theory: Associated,
color and color-acoustic metrics for the Wu-Yang monopole model},
 Phys. Rev. {\bf D 76}, 124011 (2007)

\bibitem{BZD2} A.B. Balakin, H. Dehnen, A.E. Zayats, {\it Non-minimal pp-wave Einstein-Yang-Mills-Higgs model:
color cross-effects induced by curvature}, Gen. Relat. Grav.
{\bf 40}, 2493 (2008)

\bibitem{BZD3}A.B. Balakin, H. Dehnen, A.E. Zayats, {\it  Effective metrics in the non-minimal
Einstein-Yang-Mills-Higgs theory}, Annals of Physics {\bf 323},
2183 (2008)

\bibitem{nu1} T.H. O'Dell,  {\it The electrodynamics of magneto-electric media} (North-Holland, Amsterdam, 1970)

\bibitem{NO} S. Nojiri, S.D. Odintsov, {\it Unified cosmic history in modified gravity: from F(R) theory to Lorentz non-invariant models},
Phys. Rept. {\bf 505}, 59 (2011)

\bibitem{ExactSolutions} H. Stephani, D. Kramer, M. MacCallum,C. Hoenselaers, E. Herlt,  {\it Exact solutions of Einstein's field
equations} (University Press, Cambridge, 2003)

\bibitem{dina} J. Wainwright, G.F.R. Ellis {\it Dynamical systems in cosmology}
(University Press, Cambridge, 1997)

\bibitem{7Y} E. Komatsu et al. {\it Seven-year Wilkinson Microwave Anisotropy Probe (WMAP) observations: Cosmological interpretation},
Astrophys. J. Suppl.  {\bf 192}, 18 (2011)

\bibitem{TurnerR} M.S. Turner, L.M. Widrow, {\it Inflation-produced, large-scale magnetic fields},
Phys. Rev. {\bf D 37}, 2743 (1988)

\bibitem{Od1} K. Bamba, S.D. Odintsov, {\it  Inflation and
late-time cosmic acceleration in non-minimal Maxwell-$F(R)$
gravity and the generation of large-scale magnetic fields},
JCAP {\bf 0804}, 024 (2008)

\bibitem{DH} I.T. Drummond, S.J. Hathrell,  {\it QED vacuum polarization in a background gravitational field and
its effect on the velocity of photons}, Phys. Rev.  {\bf D 22}, 343(1980)


\end{thebibliography}
\end{document}